\documentclass[mathleft
]{an}
\usepackage{graphicx}
\usepackage{rotating}
\usepackage{lscape}
\usepackage{longtable}
\usepackage{times}
\usepackage{fge}
\begin{document}

\Pagespan{789}{}
\Yearpublication{2011}%
\Yearsubmission{2010}%
\Month{11}%
\Volume{999}%
\Issue{88}%

\sloppy

\title{Newly found sunspot observations by Peter Becker from Rostock for 1708, 1709, and 1710}

\author{R. Neuh\"auser\inst{1} \thanks{Corresponding author: \email{rne@astro.uni-jena.de}}
\and R. Arlt\inst{2}
\and E. Pfitzner\inst{3}
\and S. Richter\inst{1}
}

\titlerunning{New sunspots 1708, 1709, 1710}
\authorrunning{Neuh\"auser et al.}

\institute{
Astrophysikalisches Institut und Universit\"ats-Sternwarte, FSU Jena,
Schillerg\"a\ss chen 2-3, D-07745 Jena, Germany (e-mail: rne@astro.uni-jena.de)
\and
Leibniz-Institut f\"ur Astrophysik Potsdam, An der Sternwarte 16, D-14482 Potsdam, Germany
\and
address
}

\received{16 July 2015}
\accepted{22 July 2015}

\keywords{sunspots - Maunder Minimum}

\abstract{We present a few newly found old sunspot observations from the years AD 1708, 1709, and 1710,
which were obtained by Peter Becker from Rostock, Germany.
For 1709, Becker gave a detailed drawing: 
he observed a sunspot group made up of two spots on Jan~5, 6, and~7,
and just one of the two spots was observed on Jan~8 and~9.
We present his drawing and his explanatory text.
We can measure the latitude and longitude of these two spots and estimate their sizes for all five days.
While the spots and groups in 1708 and the spot on four of the five days in January 1709 
were known before from other observers (e.g. Hoyt \& Schatten 1998),
the location of the spots in early January 1709 were not known before, so that they can now be considered
in reconstructed butterfly diagrams.
The sunspots detected by Becker on 1709 Jan~5 and 1710 Sep~10 were not known before at all,
as the only observer known for those two dates, La Hire, did not detect that spot (group).
We estimate new group sunspot numbers for the relevant days, months, and years.
The time around 1708--1710 is important,
because it documents the recovery of solar activity 
towards the end of the Maunder Grand Minimum.
We also show two new spot observations from G. Kirch for 1708 Sep 13 \& 14 as described in his letter to Wurzelbaur (dated Berlin AD 1708 Dec 19).
}

\maketitle

\section{Introduction}

The study of solar activity is important to understand the internal physics
of the Sun and (sun-like) stars as well as to possibly predict future solar activity
and space weather. The most commonly used proxy for solar activity are sunspots,
which have been observed by the unaided eye for about two millenia 
even before the invention of the telescope.

The daily Wolf or Z\"urich sunspot number R$_{\rm Z}$ for an individual observer is defined as follows: 
\begin{equation}
R_{Z} = k \cdot (10 \cdot g + n)
\end{equation}
with the total number of individual sunspots $n$, the number of sunspot groups $g$,
and the individual current correction factor $k$ of the respective observer. The
International Sunspot Number is obtained by a sophisticated calibration and average
of all contributing observers.

Hoyt \& Schatten (1998, henceforth HS98) have then defined the daily 
{\em group sunspot number} R$_{\rm G}$ as follows:
\begin{equation}
R_{G} = \frac{12.08}{N} \cdot \Sigma~~k_{i}^{\prime} \cdot G_{i}
\end{equation}
with the correction factor $k_{i}^{\prime}$ and group sunspot number G$_{i}$
of the {\em i}-th observer, and N being the number of observers used for the
daily mean (only observers with $k_{i}^{\prime}$ between 0.6 and 1.4 were used
for the time since 1848).
Without the factor 12.08, the value obtained from Equ. (2) would
be the {\em group number} instead of the {\em group sunspot number}.

In HS98, the exact dates of telescopic 
sunspot observations 
are listed 
together with the name of the observer, the place, and
the number of sunspot groups observed by that observer on each day.\footnote{See
ftp://ftp.ngdc.noaa.gov/STP/space-weather/solar-data/solar-indices/sunspot-numbers/group/daily-input-data/}
If an observer detected only one spot (or few small spots close to each other), 
HS98 consider them as one sunspot {\em group}.

Some additional telescopic sunspot observations in or around the Maunder Minimum were found after 1998, 
see publications by Vaquero (2003), Vaquero et al. (2007, 2011),
Vaquero \& Trigo (2014), Carrasco et al. (2015), and Gomez \& Vaquero (2015).
Those new data revised the daily, monthly, and yearly group sunspot numbers from HS98.

Here, we present newly found old sunspot observations for 1708, 1709, and 1710 by the
observer Peter Becker from Rostock, Germany, which also change the 
daily, monthly, and yearly group sunspot numbers compared to HS98.
The period 1708 to 1710 is of particular importance and relevance,
because it is during the last few years of the Maunder Grand Minimum.
The duration of the Maunder Grand Minimum has been discussed since
quite some time until most recently (Sp\"orer 1887, Maunder 1890, Eddy 1976,
Ribes \& Nesme-Ribes 1993, Vaquero et al. 2011, Vaquero 2012, Vaquero \& Trigo 2014, 2015, 
Usoskin et al. 2007, and Zolotova \& Ponyavin 2015).

Only with complete and precise data on sunspots in those years,
including the latitude of spots, one can date and study the end of
the Maunder Minimum accurately. The data by Becker are particulary valuable,
because his drawing of the full solar disk with the location of the spots 
is available for 1709 Jan~5--9.

In Sect.~2, we introduce the observer Becker,
we then present his German text (with our English translation) and his drawing in Sect.~3.
These observations were first reported (in German language) by 
one of us (Pfitzner 2014).
We measure the spot positions and areas in Sect.~4.
We compare the reports by Becker with other the observers for the relevant days
and calculate new group sunspot numbers for the relevant years, months, and days (Sect.~5).
We summarize our findings in Sect.~6.

\section{Observer Peter Becker from Rostock}

Peter Becker was born 1672 Nov 3 in Rostock, Germany.
He went to school in Rostock and Anklam, both in northern Germany,
and then started to study theology and mathematics at the University or Academy
of Rostock in 1690 to finish with the grade {\em Magister lib. art.} in 1696.
In 1697, he became professor for lower mathematics at U Rostock, where he tought until 1752.
He was elected rector of the university in both 1715/16 and 1723/24.
In 1714, he became Archdiakon of the local Sanct Jacobi church, 1721 its pastor,
and he worked there until his death on 1753 Nov 25.
See Fig. 1 for a painting of Becker.

Peter Becker built an astronomical observatory on the roof top of the pastor's house
of Sanct Jacobi, where he lived, at the corner of {\em Enge Stra\ss e} and {\em Lange Stra\ss e} in Rostock.
He bought and built a few instruments including a tubum terrestrem, two 12 and 20 feet telescopes,
glasses, a clock precise to the second, and a device to project the Sun onto a screen inside
a dark room.

\begin{figure}
\includegraphics[width=0.48\textwidth]{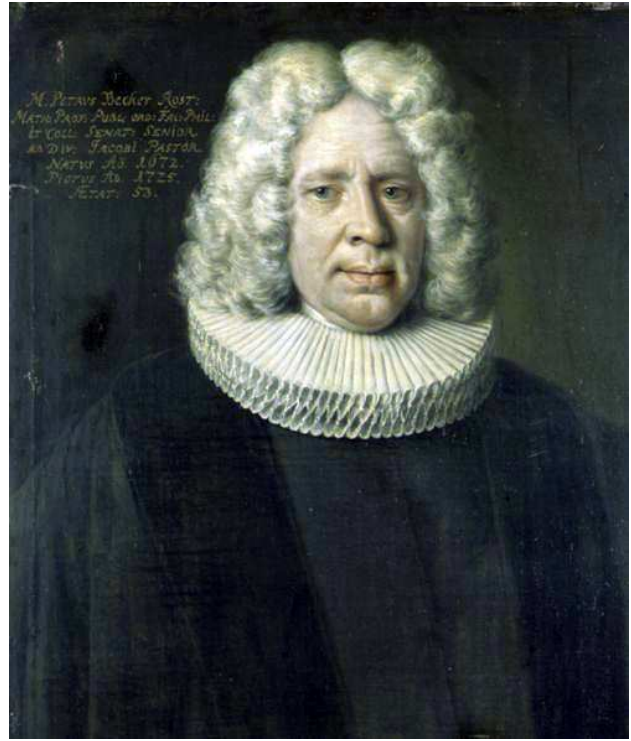}
\caption{{\bf Peter Becker.} This painting shows the observer Peter Becker.
Source: University archive U Rostock ({\em Professorenbilder}).}
\end{figure}

One of his first known astronomical observations was the solar eclipse on 1699 Sep 13,
for which he reported seven phases. 
In his work of 1703, he also mentioned sunspots.
In 1708, he was then asked by Duke Friedrich Wilhelm von Mecklenburg-Schwerin
to prepare and write the calendar for his local state of Mecklenburg-Schwerin for the years 1709 to 1715.
Only in the two calendars for 1709 and 1710, there are detailed astronomical observations listed,
while the work for the other years contain computed times of occultations and of the start of the seasons --
all for the geographic latitude of Rostock being $54^{\circ}10^{\prime}$.
Then, he published a detailed description of the Mercury transit of 1736 Nov 11 with colour figures.
Becker also wrote theological and mathematical papers.

It is quite likely that Becker used the Gregorian calendar in all his remarks on the sunspots
in the years 1708, 1709, and 1710, 
even though he himself was a protestant pastor and even though his protestant
northern German state (Dutchy) of Mecklenburg-Schwerin implemented the Gregorian reform late:
The date given by him for the solar eclipse observed by himself in 1699 was
1699 Sep 13, while the eclipse was on 1699 Sep 23 on the Gregorian calendar
(eclipse.gsfc.nasa.gov); hence, for 1699, he used the Julian calendar.
Then, for the solar eclipse (mentioned below within his report on sunspots
for 1708) was dated by him to 1708 Sep 14, which is the correct date on the 
Gregorian calendar (eclipse.gsfc.nasa.gov); hence, for 1708, he used the Gregorian calendar.
Indeed, most Central and Northern European protestant states of the former Roman Empire
implemented the Gregorian calendar in 1700 (e.g. von den Brincken 2000). 
Also, the sunspot data as given by Becker (see Table 3 below for 1708--1710)
compare well with the observations from other observers as tabulated in HS98,
who give all dates in the Gregorian style.

More details can be found in the following sources, which we consulted
at the university archive (UA) or library (UB) of U Rostock or the state library in Schwerin:
\begin{itemize}
\item UA Rostock, Personalakte P. Becker 1696-1753, Prof. Math. (employee file),
\item UA Rostock, Archivierte Sammlung 304, Gelehrtenfamilie Becker (documents of Becker family),
\item UA Rostock, Handschriftenabteilung Mss. Meckl. 0 42, Mss. math. phys. 41, Mss. math. phys. 53 (hand-written documents),
\item Landesbibliothek Schwerin, Mkl e I 75, Kalender 1709 (calendar 1709),
\item UB Rostock, Sondersammlungen, Mk-418 or LB V 23, Kalender 1710-1715 (calendars 1710-1715),
\item UB Rostock, Handschriften, Gelehrte Nachrichten auf das Jahr 1754, Rostock und Wi\ss mar, 8. Beylage, page 403, 9. Beilage pages 450-453, Sign. Z-o3 
(hand-written news from the year 1754).
\end{itemize}

\section{Observations by Peter Becker}

We will now present the reports from the observer Peter Becker for the three years,
in which he observed and reported spots.

\subsection{1708}

For 1708, the available information is summarized as follows: 
\begin{quotation}
1708: Observationes Macularum Solarium d. 10 Sep \& 1 Dec,
\end{quotation}
i.e. 
\begin{quotation}
1708: observations of sunspots on Sep 10 and Dec 1,
\end{quotation}
as specified in {\em Familienpapiere. Programma Memoriae Viti Petri Beckeri, 7. April 1754}
from page~18 onwards -- located at the state library Schwerin, Germany. 

The full text is as follows: 
\begin{quotation}
Von den Flecken in der Sonnen. \\
Ob man zwar hiesigen Ohrtes lange Zeit her vergeblich sich bem\"uhet einige Maculas oder Flecken in der Sonnen mit andern/ 
so davon vielmahls Bericht eingesandt/ zu observiren; hat man doch in dem abgelauffenen Jahr zu zweyen mahlen dazu 
erw\"unschte Gelegenheit gefunden. Und zwar anfangs den 10. Sept. da man bey Zur\"ustung der Observation der damahls 
vorstehenden Sonnen Finsterni\ss / den discum Solis auff eine Tafel in die 
cameram obscuram fallen lassen; Da 
sich denn an der Westlichen Seite des disci solaris unweit der Linie, dadurch man die eclipticam im disco abbildet/ 
etwa zwey Zoll vom Rande ein kleiner schwartzer Flecken Morgens um 10. Uhr gezeiget/ 
welcher auch noch den folgenden Tag um 11 U. 16 etwa 1 1/2  Zoll vom Rande/ und 
etwas mehr von der ecliptica declinirend/ observiret worden. Folgende Tage aber hat der tr\"ube 
Himmel nicht weiter verg\"onnet etwas von der Sonnen/ folglich auch von dieser Macula zu sehen. 
Wiewohl von Berlin her will versichert werden/ da\ss~der Flecken noch den 14./ als am Tage der Finsterni\ss / 
in der Sonnen gestanden/ und bemercket worden sey. Nachhin hat man an dem 1. Decembr. vorigen Jahres um 12. 
Uhr abermahls an der S\"udlichen Seiten unweit der Ecliptica zwischen den 3. und 4ten Zoll vom Rande 3. kleine Maculas, 
deren zwo in gerader Linie, etwa einen Zoll von der Sonnen breite von einander/ die dritte und kleinste aber 
etwas \"uber und n\"aher dem Flecken/ der nach dem Rande zu stunde/ sich gezeiget. 
Waren alle drey nicht eben sonderlich schwartz/ doch die \"oberste und kleineste am wenigsten/ 
daher man auch sonderlich scharff sehen m\"ussen/ wo man die 3te bemercken wollen. 
Weilen nun in folgenden Tagen der Himmel tr\"ube/ und innmittelst dieses zum Druck befordert werden 
m\"ussen/ soll so wohl von diesen/ als was weiter f\"urfallen m\"ochte/ k\"unfftig geliebtes Gott Bericht 
ertheilet/ und/ wo es der M\"uhe wehrt/ dem Leser in einem Kupffer alles vor Augen geleget werden.
\end{quotation}

We translate this text to English as follows staying close to the original German wording: 
\begin{quotation}
On the spots on the Sun: \\
While one was trying here for a long time without success to observe some spots on the Sun
together with others, / who have reported about them often, / there were only two cases in the
previous year with the desired opportunity. Namely first on September 10, while one was preparing
the observation for the expected solar eclipse / one has projected the solar disc on a screen
of the camera obscura. Then, on the western side of the disc of the Sun not far from the line,
which shows the ecliptic on the disc, / about two twelfths\footnote{The German word ``Zoll'' (=~inch)
was used in the sense of a twelfth of the solar diameter here, whence the usage of ``twelfth''.}
away from the edge, there was a small black spot
in the morning at 10 o'clock / which was also observed on the next day at 11:16h 
about one and a half twelfths away from the edge / a bit more inclined from the ecliptic. /  
The next days, the overcast sky did not allow to see the Sun / nor that spot.
However, it was ensured from Berlin / that the spot was still there on the 14th, /
the day of the eclipse, / on the Sun / and it was noticed. Afterwards, on the 1st of December
of last year at 12 o'clock, there were -- again on the southern side close to the ecliptic
between the three and four twelfths from the edge -- three small spots seen, two of them on a straight line,
separated by about one twelfth in solar latitude, / the third and smallest, however, was a bit closer and
above the spot, / which stood closer to the edge. /    
While all three were not very black, / but the one, which was on the top and the smallest, was
least black, / so that one had to observe very sharp, / to notice the third. 
While the sky wss overcast in the next days, / and while this had to be submitted
to press, / it should be told about this / and whatever else will happen, /
by the beloved god to be reported / and / where the effort is worth, / to show everything
to the readers eyes in a copperplate.
\end{quotation}

The source for the report is {\em Verbesserter Astronom- und Physicalischer Mecklenburgischer 
Calender Auff das 1709. Jahr nach Christi} 
(i.e. {\em improved astronomical and physical calendar for Mecklenburg for the year 1709 after Christ})
located at state library Schwerin (Disk 711a x, CD-R 80).

\begin{table*}
\caption{{\bf New sunspots.} We summarize here the details about the sunspots observed
and/or reported by Becker -- with the date and local time at Rostock 
(geographic latitude and longitude being $54^{\circ}10^{\prime}$ North and $12^{\circ}8^{\prime}$ East, respectively), 
the number of spots and groups, the hemisphere, and other details given.
We also include dates, for which the sky was reported to be overcast.}
\begin{tabular}{llllll} \hline
Date (local time)      & Spots   & Groups & Hemisphere   & Details & Observer \\ \hline
1708 Sep 10, 10h       & 1       & 1      & (south) west & {\em not far from ... ecliptic ... 2 twelfths from edge} & Becker \\
1708 Sep 11, 11:16h    & 1       & 1      & (same)       & {\em 1.5 twelfths from egde} & Becker \\
1708 Sep 12 onwards    &         &        &              & sky overcast at Rostock & Becker \\
1708 Sep 14            & 1       & 1      &              & {\em the spot was still there} & anonymous, Berlin\\ \hline
1708 Dec 1 12h noon    & 3 small & 1      & {\em again} south & {\em close to ecliptic ... 3--4 twelfths from edge} & Becker \\
1708 Dec 2 onwards     &         &        &             & sky overcast at Rostock & Becker \\ \hline
1709 Jan 3             & 0       & 0      &             & Sun ... free of even the smallest spots & Becker \\
1709 Jan 4             &         &        &             & sky overcast at Rostock & Becker \\
1709 Jan 5, 11:10h     & 2       & 1      & south Fig. 2 & {\em size of a shilling} & Becker \\ 
1709 Jan 6, 10:20-12h  & 2       & 1      & south Fig. 2 & & Becker \\
1709 Jan 7, 9:12h      & 2       & 1      & south Fig. 2 & & Becker \\
1709 Jan 8, 13:49h     & 1       & 1      & south Fig. 2 & & Becker \\
1709 Jan 9, 13:40h     & 1       & 1      & south Fig. 2 & {\em 1 twelfth from western edge} & Becker \\
1709 Jan 10 onwards    & 0       & 0      &              & {\em pretty nice sky ... but not any trace} & Becker \\ \hline
1710 Sep 10            & 1       &        &              &  & Becker \\ \hline
\end{tabular}
\end{table*}

In summary, we can conclude that there was one spot on AD 1708 Sep 10 and 11
on the western side close to the edge and close to the ecliptic;
since the solar rotation axis is almost perpendicular to the ecliptic in
early September, the spot was also close to the solar equator.
Then, there was bad weather the next days at Becker$^{ \prime}$s site, 
he was told that the spot was still seen from Berlin on Sep 14.
Also, on AD 1708 Dec 1, there were three small spots -- {\em again} on the southern side,
one twelfth off the equator and 3--4 twelfths from the edge -- apparently forming one group;
again bad weather the next days. Given that Becker uses {\em again} (German: {\em abermals})
when reporting the southern latitude of the spots in December, we may conclude that
the spot in September was also on the southern hemisphere,
as most spots in the Maunder Grand Minimum (Ribes \& Nesme-Robes 1993).

\subsection{1709}

\begin{figure*}
\includegraphics[width=\textwidth]{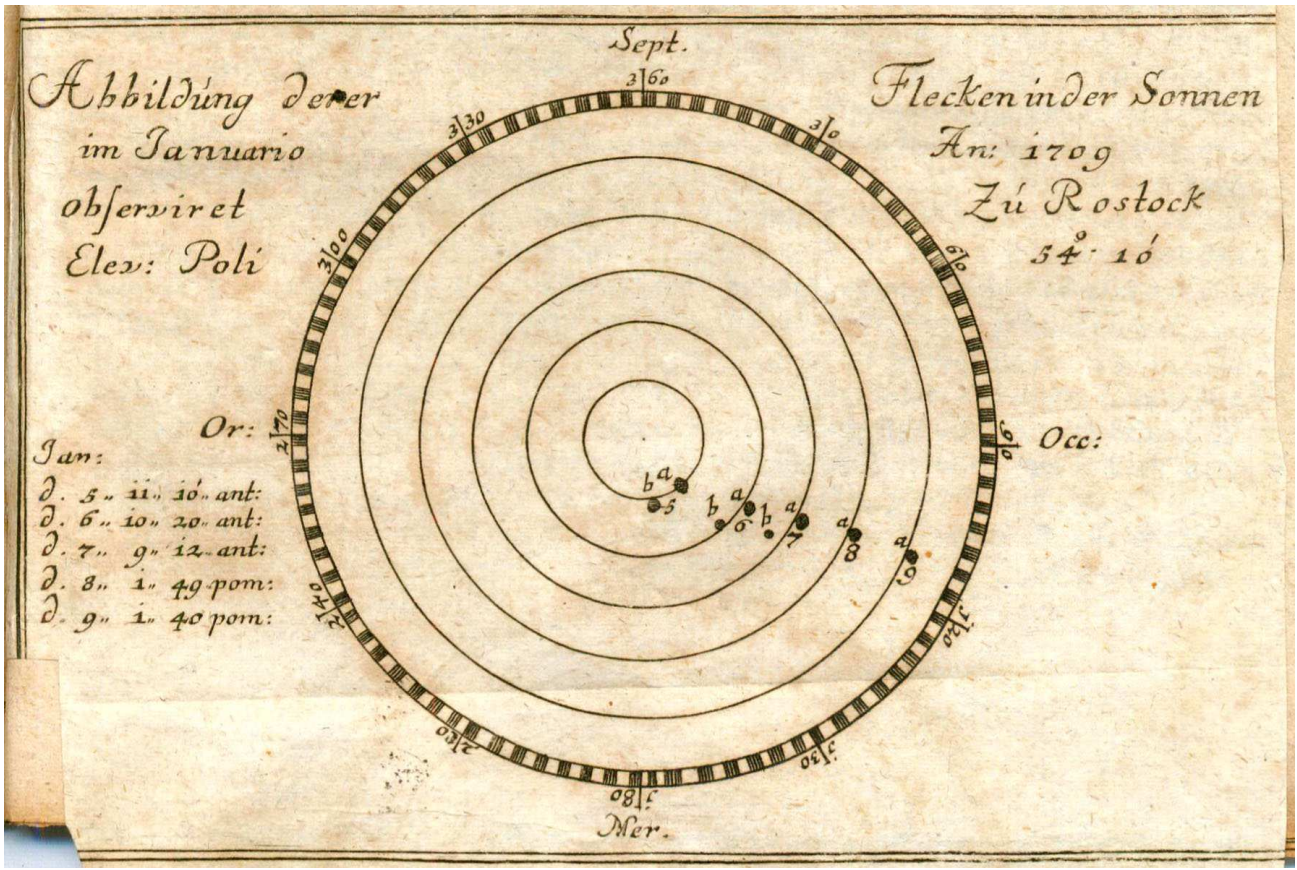}
\caption{The sunspot drawing by Peter Becker for 1709 Jan~5, 6, and~7 (one group with two spots), 
and just one of the two spots was left on Jan~8 and~9. The spots are marked {\em a} and {\em b}
and are referred to by these two letters in his text. Becker has clearly labelled the figure
with diretions (abbreviated Latin words): {\em or} for east, {\em sept} for north, {\em occ} for west,
and {\em mer} for south. He has also devided the solar disc into six sub-disks
and 12~angular parts. Then, at the top left and right, he wrote:
{\em Abbildung derer Flecken in der Sonnen im Januario An: 1709 observiret zu Rostock
Elev: Poli $54^{\circ}10^{\prime}$}, which means:
{\em Figure of the spots on the Sun in January 1709 as observed in Rostock at
latitude $54^{\circ}10^{\prime}$.}
On the left, the observing times (local times Rostock) are given: 
{\em Jan~5 11:10 am, (Jan)~6 10:20 am, (Jan)~7 9:12 am, (Jan)~8 1:49 pm, and (Jan)~9 1:40 pm}.
\label{1709_01_05}}
\end{figure*}

The calendar (diary) from 1710 by Peter Becker includes not only the text about the
sunspots (see below) as observed in 1709, but also a copperplate with a drawing of the spots.
The plate has the following caption: 
\begin{quotation}
Abbildung derer Flecken in der Sonnen im Januario An: 1709 observiret zu Rostock,
\end{quotation}
which we translate to English as follows: 
\begin{quotation}
Figure of spots on the Sun in January of the year 1709 observed at Rostock.
\end{quotation}
This calendar is available as {\em Verbesserter Calender 1710} by Becker in the university libraries
of U Rostock and U Schwerin.
We show the original drawing of the sunspots from 1709 by Peter Becker,
as observed from Rostock, Germany, in Fig.~\ref{1709_01_05}.

The observer Becker has written the following explanatory text about his observations in 1709: 
\begin{quotation}
Von den Observationibus des vorigen 1709ten Jahres \\
Zweierley Merckw\"urdigkeiten sind es aus vorigem Jahr / die dem geneigten Leser mitzutheilen vor
n\"othig erachtet habe / von dem / was in Astronomicis von uns ist observiret worden. Das erste
zwar / der Zeit nach / sind die merkliche Sonnenflecken / so in dem Januario 1709. sich darinn / zu
nicht geringer Vergn\"ugung derer Anschauenden/ zeigeten. Ich habe mich bem\"uhet, die gantze
observation in einem Kupfer Stich vor Augen zu legen / welches wer es anzusehen die M\"uhe
nehmen wird / kan gar leicht = und deutlich erkennen / so wohl die Gr\"osse als Stelle der Flecken /
wie sie an jedem Tage erschienen. Und damit der G. Leser den gantzen Umstand der Observation
deutlich wisse; wird er mercken / da\ss~ich die Sonne durch einen Tubum etwa 4. Schuh lang in einen
duncklen Ort / mit H\"ulfe eines oculi versatilis hinein fallen lasse / und auff einer 
gegen \"uber stehenden und ad perpendiculum immer gerichteten Tafel (daran ein wei\ss~Pappier 
befestiget) auf fange.
Wann ich nun den wahren diametrum der Sonnen / nach der einmahl festgestelten
distance der Tafel von dem oculair-Glase des Tubi, mit Flei\ss~genommen / und darnach den discum
lis auffgeri\ss en / und in 12. digitos 6. paralel Zirckel getheilt; so notire ich die Stelle und Gr\"o\ss e der
Flecken / mit Anmerckung der Zeit / so wohl nach einen Uhr das minuten und secunden zeiget / 
al\ss~sciaterico, und H\"ohe der Sonnen. Anlangend nun den Abri\ss~/ ist zu mercken / wenn man das Kupfer
nach dem beygeschriebenen Gegenden der Welt Or. Occ. \& c. recht h\"alt / da\ss~es als denn die Sonne
also vorstelle / als sie am Himmel mit 
Angesehen worden. Bei den Flecken selber aber / ist wohl zu notiren / 
da\ss~sie vermuthlich pl\"otzlich in der Mitte der Sonnen entstanden. Welches daraus erhellet / weil an dem 
dritten Jan. die Sonne gantz rein und sonder den geringsten Flecken gewesen. Am 4. Jan. war der
Himmel tr\"ube/ und verg\"onnte die Sonne nicht zu sehen. Am 5. aber um 11.h 10. zeigeten sich 2
ziemlich merckliche Flecken wie a.b. mit der Zahl 5. in der Figur weiset, So da\ss~sie durch einen
Tubum von 7. Schuh in einer distance von 8 Schuh wohl wie ein Schilling von Gr\"o\ss e sich zeigeten.
Wie wohl nicht zu leugnen / daß der Flecken b an dem Tage weniger schwarz erschiene als der
andere a. Am folgenden 6. Tage um 10 Uhr 20 Min. war der eine b. gar kleiner / und daher wenig zu
sehen / so da\ss~ich fr\"uh morgens bei aufgehender Sonne 
ihn gar nicht erblickte / 
wie wohl er gegen Mittag im nun deutlicher Wand. \\
Am 7. um 9. Uhr 12. Min. waren beide wieder kenntlich und schwartz / allein der in a viel gr\"o\ss er als
der andere in b. Am 8ten um 1h49. pom. war b gar hinweg / und nur der in a alleine befindlich; wie
auch am 9ten um 1.Uhr 40. Min der eine in a nur allein / und noch einen Zollbreit vom Westlichen
Rande der Sonnen. Da denn weil den gantzen Tag durch ein heiteren sch\"onen Himmel / ich mit allen
Flei\ss~nach dem Flecken b. gesehen / aber auf keine Art nur die geringste Spuhr 
davon erblicken m\"ogen. \\
Von diesen Flecken nun meine Gedanken zu er\"offnen enthalte mich / eile aber vor dieses mahl /
und begn\"uge mich dem G. Leser die Observationen deutlich vorgestellt zu haben.
\end{quotation}

We translate this text to English as follows --  again staying close to the original German wording: 
\begin{quotation}
About the observations of the previous 1709th year: \\
There were two curiosities in the previous year, / 
which need to be reported to the inclined reader, / about that /
what has been observed by us in astronomy. The first, however, / regarding the time, /
were the remarkable sunspots / like in January 1709. They were seen with
not that small joy by the observers. I have tried to present the whole observation
on the cupper plate, / so that those, who will take the effort to view it, 
can pretty easy and clearly recognize them, / as well as the size and the location
of the spots, / as they were seen on those days. And so that the inclined reader
know well the whole circumstances of the observation; he will notice, / that 
I let the Sun fall through a tube, about four feet long, at a dark place, /
with the help of a movable eyepiece / and I then collect it
on a perpendicular screen on the opposite side (on which a white paper is fixed).
When I now have taken with effort the true diameter of the Sun /
using the once fixed separation of the screen from the ocular glass of the tubus, /
and then drawn the disc of the sun, / and devided into 12 positional angles and 6 parallel circles;
then I mark the location and size of the spots / with a notice of the time /
both with a clock that shows hours and minutes / also with a shaddow stick,
and from the height of the Sun. Regarding the drawing / it is to be remarked, /
when holding the cupper plate relative to the areas of the world towards east, west etc., /
that one can see the Sun / as seen on the sky.
For the spots themselves, it is to be noticed, / that they presumably formed
suddenly in the midth of the Sun. This is concluded from the fact, / 
that the Sun was fully clean on the 3rd of January and free of even the smallest spots.
On the 4th of January, the sky was overcast, so that the Sun was not visible.
But on the 5th at 11:10h, two rather remarkable spots were seen, which are marked
a.b. with the number 5 in the figure. They were seen through a tube of seven feet
at a separation of eight feet, at a size of a shilling.
As it can not be neglected, / that the spot b was less dark on that day then spot a.
On the following 6th day at 10 hours and 20 minutes, the one marked b was rather smaller /
and therefore less well visible, / so that I could not see it early in the morning
after Sun rise / while it was clearly seen around midday. \\
On the 7th at 9 hours 12 minutes, both were again clear and black /
but the spot a was much larger than the other spot b. On the 8th at 1:49h after midday,
b was gone / and now only a was there; also on the 9th at 1 hour 40 minutes,
only a was there alone / and still one twelfth away from the western edge of the Sun.
While throughout all the next days, we had a pretty nice sky, / I looked for the spot b.
with all effort, / but I could not notice any trace of it anymore. \\
To inform about my own thoughts about those spots, I refrain, /
but I hurry up / this time, / and 
it is sufficient to have 
presented these observations clearly to the inclined reader.
\end{quotation}

In summary, Becker has seen and drawn a spot group with two spots for three days,
and one of them for two more days, as shown in Fig.~\ref{1709_01_05}.
Becker clearly explains that he tried to present size and position of the spots well
and close to reality. He also gives the time of observations precisely to the minute.
The spots were close to the western edge of the Sun, and also close to the equator
(on the southern hemisphere, as seen from the figure). He reported the size of the spots
to be {\em a schilling} when seen through a {\em tube of seven feet at a separation of eight feet}.

\subsection{1710}

For 1710, the available information is just:
\begin{quotation}
{\em 1710: Observationes Macularum Solarium\\ d. 10 Sep}
\end{quotation}
i.e. 
\begin{quotation}
{\em 1710: observations of sun spots on Sep 10}.
\end{quotation}
We can conclude here only that (at least) one sunspot (group) was observed on AD~1710 Sep~10.

\section{Sunspot latitude and size}

The verbal descriptions of spot locations in 1708 are not very 
precise but good enough for estimates of the heliographic 
positions. Figure~\ref{1708_09_10} shows a reconstruction of
the location of the spot on the solar disk with a heliographic
coordinate system. The main information is the distance from the
western solar limb, while the statement `near the ecliptic' is
hard to interpret. Since Becker later claimed that both the
Sep~10 and the Dec~1 spots were on the southern side, we
assume that the spot was actually on the solar equator which 
is south of the ecliptic in September, but not too far away
from it to be incompatible with being `near'.
The spot is then not deviating from the ecliptic more than
on the day before, but we consider this reasonably good given
the limited accuracy of such textual descriptions. Alternatively,
we may interpret the text as `a bit more [than those 1.5~twelfths
from the limb] inclined from the ecliptic.' However, this
would place the spot near $-15\degr$ heliographic latitude
and is incompatible with the other reports below.

Sp\"orer (1889) mentioned that 
Wideburg (also called Wiedeburg, while HS98 gave Wiedenburg)
had observed 
small spots 1708 Sep 2--14 (in our Tab. 3 for Sep 10 \& 11) at a heliographic latitude of $-5^{\circ}$,
consistent with Becker's text for presumably the same spot.
In Ribes \& Nesme-Ribes (1993), we can find four data points
in September 1708 at heliograhpic latitudes of 
$-4^{\circ}$, $-5^{\circ}$, $-6^{\circ}$, and $-8^{\circ}$,
possibly from La Hire (who detected one spot (group) on 
13 days between 1708 Sep 3-18 according to HS98, see also Tab. 3 below),
also consistent with Becker's text for that spot.
 
G. Kirch describes his spot observations for AD 1708 Sep 13 
in a letter to Wurzelbaur (dated Berlin AD 1708 Dec 19) as follows
(cited in German from the Kirch letter compilation in Herbst 2006):
\begin{quotation}
[15:25h] war die Macul 8 partes micrometri 10 schuhigen Tubi vom West-Rande, ist $1^{\prime}$ $9^{\prime \prime}$ Diameter Solis war 224 p.m. ist $32^{\prime}$.
\end{quotation}
We translate this to English as follows:
\begin{quotation}
[15:25h] in the 10-foot tube, the spot was 8 mirometer parts (p.m.) away from the western egde, 
which is $1^{\prime}$ $9^{\prime \prime}$, the diameter of the sun was 224 p.m., which is $32^{\prime}$.
\end{quotation}
This was given for 3:25h p.m., presumable his local mean solar time.
G. Kirch also observed the solar eclipse the next day and wrote in the same letter:
\begin{quotation}
[...] das beste bey der Sonnenfinsternis [...]: nemlich die sehr kleine Macul, 
welche nahe am West-Rande der Sonnen noch zu erblicken war. 
Diese ward nicht vom Mond bedeckt, sondern da man sie am n\"achsten beym Mond zu seyn sch\"atzete, 
war sie etwan 6 p.m. 10 schuh. Tubi von ihm, 
w\"are $51^{\prime \prime}$. und dieses war um 8 Uhr 8 min, 41 sec. 
Man konte sie nicht stets erkennen, weil sie sehr schwach 
und klein war, und der Himmel auch ein mal reiner war, als das andere.
\end{quotation}
We translate this to English as follows:
\begin{quotation}
... the best during the solar eclipse was the very small spot, which was visible close to the western edge on the sun.
This part was not eclipsed by the moon, it was also quite close to the lunar limb,
so it could be estimated to lie about 6 p.m. [mirometer parts in the] 10-foot tube from it [the lunar limb],
i.e. $51^{\prime \prime}$. And this was at 8 o'clock 8 min 41 sec. One could not always see it,
because it was very weak and small, and the clearness of the sky was variable.
\end{quotation}
G. Kirch also wrote that he observed the solar eclipse 
on AD 1708 Sep 14 from 7:31h (estimated by him) to 9:42h (measured by him), in the morning,
presumably his own local mean solar time (in Berlin, Germany). 
Given the eastern longitude of the old Berlin observatory, this corresponds to 6:37.5 to 8:48.5h UT.
We estimated with StarCalc that the eclipse lasted from UT 6:38 to 8:40h UT
at his location (Berlin observatory $52^{\circ} 31^{\prime} 8^{\prime\prime}$ north, 
$13^{\circ} 23^{\prime} 29^{\prime\prime}$ east in Berlin-Dorothenstadt), 
which gives a time difference to G. Kirch's clock of 0 to 8 min.
We apply the same difference to his sunspot observation (8:08h local time given by G. Kirch), 
i.e. shifting by 4 to 8 min. His descriptions for the spot on Sep 13 and 14 are basically: 
(a) seen around 3h 25 min in the afternoon (his local time) at the western edge (solar diameter 224 p.m.) for Sep 13, 
and (b) close to the western edge, not covered by the moon, 6 p.m. from the lunar limb, 
not always visible (for Sep 14 at 8h and 8 min in the morning).

G. Kirch's description of the spot position at the specified time allows us to reconstruct it using StarCalc and Kirch's data
(for the geographic position of the old Berlin observatory as given above), see Fig. 3.
Because we determined a time difference of Kirch's clock of 0 to 8 min, 
we recalculated the spot's position for several times to determine the error bar. 
We conclude on a heliographic latitude of the spot of $-4.5 \pm 1.5^{\circ}$. 
This is fully consistent with the values from Becker (our Tab. 2), from Sp\"orer for Wideburg, 
and in Ribes \& Nesme-Ribes (1993) for La Hire.

\begin{figure}
{\includegraphics[width=0.48\textwidth]{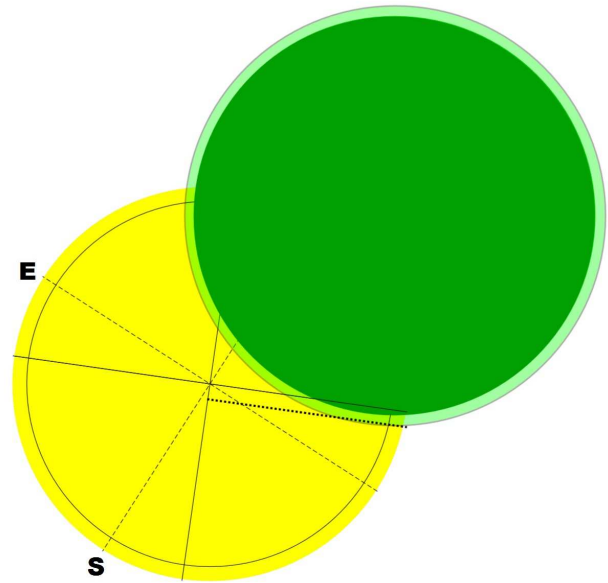}}
\caption{Reconstruction of the location of the spot as observed by G. Kirch on AD 1708 Sep 14
during a solar eclipse: we see the moon in the upper right and the Sun in the lower left (StarCalc).
The inner ring in the sun indicates the separation to the western limb as given by G. Kirch for 
the spot on Sep 13 (8 micrometer parts, p.m.), and the outermost ring around the moon indicates 
the spot's separation to the lunar limb on Sep 14 (6 p.m.). 
The dashed lines mark the directions East (E, left), South (S, bottom), etc.
The full nearly horizontal line is the sun's equator -- nearly identical to the ecliptic and directed from ESE to WNW. 
The dotted line parallel to the sun's equator indicates the best heliographic latitude for the spot
slightly south of the solar equator: 
because the separation of 8 p.m. was measured one day after the eclipse, 
the real position of the spot on Sep 14 lies near the intersection of the inner circle in the sun
and the circle outside the moon. }
\end{figure}

We would also like to note that this particular spot observation by G. Kirch was not listed in HS98,
who list only one spot on AD 1708 Sep 11 for G. Kirch (HS98), 
otherwise they give {\em not observed} for G. Kirch for that month.
In his letter to Wurzelbaur (see Herbst 2006), G. Kirch mentioned that he observed 
a spot (group) on AD 1708 Sep 2, 3, 13, and 14.
Since we discuss his spot observations only for comparison with Becker, we will discuss only Sep 13 \& 14 here,
while we will report all new spot observations found for G. Kirch in his letters 
(Herbst 2006) elsewhere (Richter et al. in prep.).

The description for 1708 Dec~1 is less conflicting; the 
corresponding spot reconstruction is shown in Fig.~\ref{1708_12_01}.
Note that the reconstructions rely on distances to the ecliptic
which can neither be easily inferred from the day time sky nor from
a particular mount of the telescope. The resulting heliographic
positions need to be treated with care. 
For AD 1708 Nov 30 and Dec 1, Sp\"orer (1889) reported a spot group as observed by Wideburg
(also in our Tab. 3 for Dec 1) at $-10^{\circ}$, consistent with our value for Becker (Tab. 2).

The drawing of the solar disk showing spots on 1709 Jan 5--9
allows us to measure positions and estimate areas. The image shows
a coordinate system which is most likely meant to be an equatorial
one. Becker refers to the horizontal axis as {\em the world's east 
and west}. If it is horizontal, the spots would not show the solar 
rotation so well due to the different observing times on the 
individual days. The same holds true for an ecliptical system as 
used by Scheiner and Hevelius for example, which also leads to a 
mismatch with the rotation (the latitude of the main spot would 
increase from $-13\degr$ to $-27\degr$). The east and west marks also
indicate that the image is not mirrored as it would be on the front
side of a projection screen. This is supported by a drawing of the
solar eclipse of 1709 Mar 11 by Becker, 
which is also not mirrored.

\begin{figure}
{\includegraphics[width=0.48\textwidth]{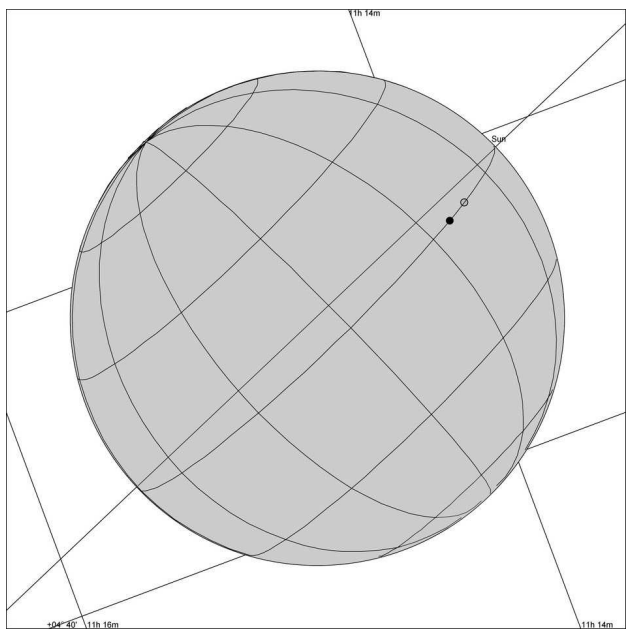}}
\caption{Reconstruction of the location of the spot described by Becker
on 1708 Sep~10 (filled circle) and Sep~11 (open circle).
\label{1708_09_10}}
\end{figure}

\begin{figure}
{\includegraphics[width=0.48\textwidth]{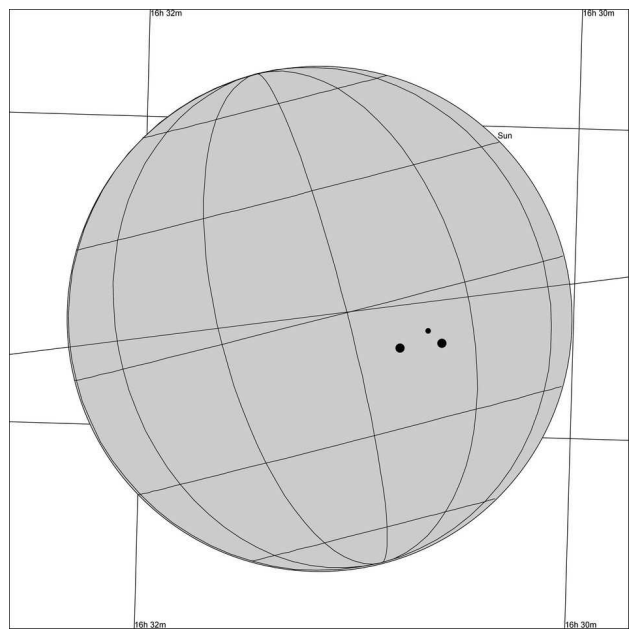}}
\caption{Reconstruction of the location of the spots described by Becker
on 1708 Dec~1.
\label{1708_12_01}}
\end{figure}

In a first attempt, we measured the sunspot positions assuming 
a correct indication of the celestial coordinate system. The 
ephemeris for the heliographic coordinates of the solar disk 
center are taken from the JPL Horizons ephemeris 
service\footnote{http://ssd.jpl.nasa.gov/horizons.cgi} using
the algorithms of 2013 July 13 and the physical properties
of the Sun as revised on 2014 Jan 16. The ephemeris also provides 
the counter-clockwise angle of the solar rotation axis measured 
from the true-of-date celestial north pole (position angle). The 
angles result in a heliographic coordinate system that can be 
superimposed to the drawing. Apart from the rotation, the disk 
center latitude and the position angle change slightly during 
the course of the five observed days.

With these assumptions, the heliographic latitude of the main 
spot in the drawing increases gradually from $-12\degr$ on 
1709 Jan~5 to $-20\degr$ on Jan~9. We believe that 
Becker did not align his coordinate system with the true
celestial one properly.

\begin{table*}
\caption{Positions of the sunspots  
observed and plotted by Becker in 1708 and 
1709. $L_0$ and $b_0$ refer to the heliographic coordinates of the
center of the solar disk at the given time, CMD is the central meridian distance, $L$ is the heliographic
longitude, $b$ is the heliographic latitude, $\delta$ is the angular 
separation from the solar disk centre, $A$ is the apparent area in
millionths of the solar hemisphere, and $A_{\rm c}$ is the area
corrected for foreshortening. The verbal descriptions of 1708 did
not allow area estimates.}
\begin{tabular}{lp{0.3cm}rrp{0.3cm}rrrp{0.3cm}rp{0.3cm}cc}
\hline
Date       &&\multicolumn{1}{c}{$L_0$}&\multicolumn{1}{c}{$b_0$}
           &&\multicolumn{1}{c}{CMD}&\multicolumn{1}{c}{$L$}&\multicolumn{1}{c}{$b$}
           &&\multicolumn{1}{c}{$\delta$} && $A$ & $A_{\rm c}$\\
\hline
1708 Sep 10 && $326\fdg0$ & $7\fdg2$ && $41\fdg1\pm4\fdg0$  &  $  7\fdg1\pm4\fdg0$ & $ (0\fdg0\pm4\fdg5$) && $41\fdg6$ && -- & --\\
1708 Sep 11 && $312\fdg2$ & $7\fdg2$ && $48\fdg8\pm4\fdg6$  &  $  1\fdg0\pm4\fdg6$ & $ (0\fdg0\pm4\fdg8$) && $49\fdg2$ && -- & --\\
\hline
1708 Dec 01 && $323\fdg5$ & $0\fdg3$ && $16\fdg7\pm3\fdg2$  &  $340\fdg2\pm3\fdg2$ & $-10\fdg6\pm3\fdg4$ && $19\fdg9$ && -- & --\\
            && $323\fdg5$ & $0\fdg3$ && $27\fdg3\pm3\fdg5$  &  $350\fdg8\pm3\fdg5$ & $-11\fdg9\pm3\fdg9$ && $29\fdg7$ && -- & --\\
            && $323\fdg5$ & $0\fdg3$ && $24\fdg3\pm3\fdg3$  &  $347\fdg8\pm3\fdg3$ & $ -8\fdg4\pm3\fdg7$ && $25\fdg7$ && -- & --\\
\hline
1709 Jan 05 && $222\fdg9$ &$-3\fdg9$ &&  $8\fdg1\pm3\fdg0$  &  $231\fdg0\pm3\fdg0$ & $-10\fdg1\pm3\fdg0$ && $ 6\fdg3$ && 63 & 63 \\
            && $222\fdg9$ &$-3\fdg9$ &&  $4\fdg5\pm3\fdg0$  &  $227\fdg4\pm3\fdg0$ & $-14\fdg7\pm3\fdg2$ && $11\fdg7$ && 47 & 48 \\
1709 Jan 06 && $210\fdg2$ &$-4\fdg0$ && $21\fdg1\pm3\fdg2$  &  $231\fdg3\pm3\fdg2$ & $-11\fdg1\pm3\fdg5$ && $22\fdg1$ && 47 & 51 \\
            && $210\fdg2$ &$-4\fdg0$ && $16\fdg9\pm3\fdg2$  &  $227\fdg1\pm3\fdg2$ & $-15\fdg1\pm3\fdg5$ && $20\fdg0$ && 38 & 40 \\
1709 Jan 07 && $197\fdg6$ &$-4\fdg1$ && $30\fdg9\pm3\fdg5$  &  $228\fdg5\pm3\fdg5$ &  $-9\fdg3\pm4\fdg0$ && $31\fdg1$ && 63 & 74\\
            && $197\fdg6$ &$-4\fdg1$ && $25\fdg6\pm3\fdg4$  &  $223\fdg2\pm3\fdg4$ & $-12\fdg9\pm3\fdg8$ && $26\fdg8$ && 25 & 28 \\
1709 Jan 08 && $181\fdg9$ &$-4\fdg2$ && $43\fdg2\pm4\fdg1$  &  $225\fdg1\pm4\fdg1$ & $-10\fdg4\pm4\fdg6$ && $43\fdg3$ && 53 & 73\\
1709 Jan 09 && $168\fdg8$ &$-4\fdg3$ && $60\fdg4\pm6\fdg0$  &  $229\fdg2\pm6\fdg0$ & $-10\fdg8\pm5\fdg3$ && $60\fdg1$ && 47 & 94\\
\hline
\end{tabular}
\end{table*}

An alternative approach provides us with sunspot positions
independently of the coordinate system Becker drew. We determined
the optimum position angle of the solar disk for which the
positions of the larger of the two spots match the solar rotation. 
We used the differential rotation derived by Balthasar et al. (1986)
based on sunspots of 1874--1976. For future applications, one 
needs to bear in mind that the derivation of the solar rotation 
from the positions obtained with this method is no longer meaningful.
The algorithm is the same as was used by Arlt et al. (2013).

Only the larger of the two spots (spot labelled {\em a} in Fig. 2) plotted by Becker was used
to determine the position angle, since it was visible on all
the five days. The model assumes a constant position for this
spot; the coordinates obtained are $L=229\fdg0$, $b=-10\fdg4$. 
While in principle, Becker might have placed his paper differently
on each day, there is too little input information (one spot only)
for the determination of five independent position angles. Since
the spots align well in the drawing, the model assumes a constant
position angle. For a fixed direction to north, the solar coordinate
grid does actually change orientation by $1\fdg9$ from Jan~5--9,
but we consider this error small compared to the plotting errors
of the spots. The position angle derived is $-13\fdg9\pm6\fdg2$
and is $11\degr$ larger than the average one that is obtained
from Becker's north direction ($-2\fdg9$). The uncertainties
indicated in Tab.~2 assume a general plotting accuracy of 
$\Delta_{\rm plot}=3\degr$ in the disk center. The longitudes 
are mostly affected by this uncertainty, corrected for the distance 
from the center $\delta$, i.e. approximately 
$\Delta_{\rm plot}/\cos\delta$. The latitudes are additionally affected 
by the uncertainty of the orientation, which is assumed to be 
$\Delta_{\rm ori}=5\degr$ which is reached at the solar limb. The total
error for the latitude is therefore approximated by 
$\sqrt{\Delta_{\rm plot}^2 + (\Delta_{\rm ori}\sin\delta)^2}$.

Sp\"orer (1889) mentioned spot observations from G. Kirch und Wideburg for AD 1709 Jan 6--10 
(also in our Tab. 3) at b=$-16^{\circ}$,
again consistent with our values for Becker (Tab. 2).

Areas are estimated with a few cursor circles on the screen. 
The ones employed here contain areas $\tilde{A}$ of 145, 
221, 270, 308, and 364 pixels, while the entire solar disk 
contained $\tilde{A}_\odot=2.88\cdot10^{6}$ pixels. The spot 
area in millionths of a solar hemisphere (MSH) is then 
$A=10^6\tilde{A}/2\tilde{A}_\odot$. The correction for 
perspective foreshortening towards the solar limb is 
$A_{\rm c}=A/\cos d$, where $d$ is the angular separation
of the spot from the solar disk center. Note that the areas
drawn by Becker are at best estimates for the total spot area
including the penumbrae. They are more likely drawn with too
large areas -- a typical consequence of observing with very 
small telescopes showing blurred sunspots. 

These sunspot observations can now be used in future butterfly diagram
reconstructions of the last Schwabe cycle at the end of the Maunder Minimum.
The last sunspot minimum marking the end of the Maunder Minimum was dated
to 1712 by HS98. The first spot clearly on the solar 
northern hemisphere occurred in AD 1705 (Ribes \& Nesme-Ribes 1993) -- 
after decades of purely southern spots.

\section{New group sunspot numbers 1708--1710}

We can now compare the observations by Becker with other observers;
and we can then use his data to improve the daily, monthly, and yearly
group sunspot numbers for the respective days, months, and years.
Data for other observers can be found
in HS98 (see Tab.~\ref{groupnumbers}).

\begin{table*}
\caption{{\bf Other sunspot observations in 1708--1710.}
List of daily, monthly, and yearly telescopic group sunspot numbers 
by other observers on the dates with information from Becker. 
All data for those other observers are from Hoyt \& Schatten (1998, HS98), from their table {\em alldata}. 
We give {\em n/o}, if it is given in HS98 that there was {\em n}o {\em o}bservation by the particular observer
(in case of Becker, this is due to overcast sky);
we give {\em n/a}, if data from that observer and that year is {\em n}ot {\em a}vailable, i.e. not listed in HS98.
Then, we list the daily, monthly, and yearly group sunspot numbers from HS98 (with error bars)
and as re-calculated by us 
(in some cases sligthly different, even though we would have left
out $2\sigma$ deviations when taking the mean as described in HS98;
our values are consistent with those in HS98 within $1\sigma$ error bars;
days without spot observations were not considered when taking the mean).
In the next column, we give the number of groups and sunspots as observed by Becker from Tabs.~1 and~2.
In the last column, we give the new mean group sunspot numbers,
where the observations by Becker are taken into account.
The third row lists the correction factors $k^{\prime}$ for the observers from HS98
and our estimate for Becker (see text).\label{groupnumbers}}
\begin{tabular}{c|ccccccc|cc|cc|c} \hline
Date  & La  & Man- & Wide-   & G.   & Blan- & Wolf & Mul-  & \multicolumn{2}{c}{mean w/o Becker}  & \multicolumn{2}{c}{Becker} & mean w/  \\
      & Hire& fredi& burg    &Kirch & chini &      & ler   & HS98 & by us & group & spot  & Becker  \\ \hline
~~~~$k^{\prime}$&0.996&1.560&1.026&1.188&0.808 & 1.070&1.000& (a) & (a,e) & $\sim 1.11$&    & (b,e)    \\ \hline\\[-1ex]
& \multicolumn{8}{c}{1708 September} & & \\ \hline
10    &  1  & 0    & 2       & n/o  & n/o   & n/a  & n/o & $21 \pm 4$    & $12.3 \pm 12.4$ & 1   & 1   & $12.6 \pm 10.1$  \\
11    &  1  & 1    & 2       & 1    & 1     & n/a  & n/o & $15 \pm 3$    & $16 \pm 6$      & 1   & 1   & $15.5 \pm 5.4$   \\
12    &  1  & 1    & n/o     & n/o  & 1     & n/a  & n/o & $12 \pm 1$    & $13.5 \pm 4.7$  & n/o & n/o & $13.5 \pm 4.7$   \\
13    & n/o & n/o  & n/o     & 1 (e) & n/o  & n/a  & n/o & $15 \pm 4$    & $14.4 \pm 0.8$  & n/a & n/a & $12.1 \pm 0.8$   \\
14    &  1  & n/o  & 0       & 1 (e) & n/o  & n/a  & 1   & $12 \pm 0$    & $9.6 \pm 6.5$   & (1)$^{\rm f}$ & (1)$^{\rm f}$ & $9.6 \pm 6.5$    \\ \hline
month & 0.6 & 0.7  & 1.7     & 1.0  & 1.0   & n/a  & 1.0 & $8.9 \pm 7.6$ & 9.6   & 1.0 &     & 9.1 \\ \hline\\[-1ex]
& \multicolumn{8}{c}{1708 December} & & & \\ \hline
1     &  1  & n/o  & 1       & 1    & n/o   & n/a  & n/o       & $12 \pm 0$ & $12.9 \pm 1.2$ & 1   & 3   & $13.0 \pm 1.0$ \\
2     & n/o & n/o  & n/o     & 1    & n/o   & n/a  & n/o       & $14 \pm 0$ & $14.4 \pm 0.8$ & n/o & n/o & $14.4 \pm 0.8$ \\ \hline
month & 0.1 & n/o  & 1.0     & 0.3  & n/o   & n/a  & n/o$^{\rm c}$ & $1.5 \pm 4.7$ & 1.6 & 1.0 &  & 1.6 \\ \hline
year  &     &      &         &      &       &      &           & $2.8 \pm 0.1$ & 2.8  &     &     & 2.8 \\ \hline \hline\\[-1ex]
& \multicolumn{8}{c}{1709 January} & & & \\ \hline
3     & n/o & n/o  & n/o     & n/o  & n/a   & n/o        & n/o & $0 \pm 0$     & n/o  & 0    & 0   & 0.0 \\
4     &  0  & n/o  & n/o     & n/o  & n/a   & n/a        & n/o & $0 \pm 0$     & 0.0  & n/o  & n/o & 0.0 \\
5     & n/o & n/o  & n/o     & n/o  & n/a   & n/a        & n/o & $0 \pm 0$     & n/o  & 1    & 2   & $13.4 \pm 3.4$ \\
6     &  1  & n/o  & 1       & 1    & n/a   & 2$^{\rm d}$& 1   & $12 \pm 1$    & $15.3 \pm 6.0$ & 1    & 2   & $14.9 \pm 5.4$ \\
7     &  1  & n/o  & 1       & 1    & n/a   & n/a        & 1   & $12 \pm 0$    & $12.6 \pm 1.2$ & 1    & 2   & $12.8 \pm 1.1$ \\
8     & n/o & n/o  & 1       & n/o  & n/a   & n/a        & 1   & $12 \pm 0$    & $12.2 \pm 0.2$ & 1    & 1   & $12.6 \pm 0.7$ \\
9     & n/o & n/o  & 1       & 1    & n/a   & n/a        & 1   & $12 \pm 0$    & $12.9 \pm 1.2$ & 1    & 1   & $13.1 \pm 1.0$ \\
10    &  1  & n/o  & 1       & 1    & n/a   & n/a        & n/o & $12 \pm 0$    & $11.8 \pm 1.4$ & 0    & 0   & $9.6 \pm 6.5$  \\ \hline
month & 0.4 & n/o  & 1.0     & 0.7  & n/a   & 2.0$^{\rm d}$&1.0& $4.3 \pm 5.8$ & 4.5  & 0.8  &     & 5.6 \\ \hline
year  &     &      &         &      &       &            &     & $1.6 \pm 0.1$ & 1.6  &      &     & 1.7 \\ \hline \hline\\[-1ex]
& \multicolumn{8}{c}{1710 September} & & & \\ \hline
10    &  0  & n/a  & n/a     & n/o  & n/a   & n/a  & n/a & $0 \pm 0$     & $0 \pm 0$ & 1   & 1 & $6.7 \pm 9.5$  \\ \hline
month & 0.0 & n/a  & n/a     & n/o  & n/a   & n/a  & n/a & $0.0 \pm 0.0$ &   0 & 1.0 &   & 0.6  \\ \hline
year  &     &      &         &      &       &      &     & $0.4 \pm 0.0$ & 0.4 &     &   & 0.4  \\ \hline\\[-1ex]
\end{tabular}

Remarks: (a) Daily mean from HS98 (table {\em dailyrg.dat} on ftp://ftp.ngdc.noaa.gov/STP/space-weather/solar-data/solar-indices/sunspot-numbers/group/).
We also list the numbers we calculated; they are sometimes different from the numbers in HS98, even though they are
supposed to be based on the same data base: 
in some cases, HS98 may have just rounded down or omitted the digits after the comma,
but in some cases the difference in larger; in the table {\em dailynum.dat},
HS98 provide the numbers of observers per day, 
e.g. for AD 1708 Sep~11, they give {\em 3~observers}, even though according to their
table {\em alldata}, La Hire, Wideburg, G.~Kirch, and Blanchini are the four observers of that day.
The largest difference between their and our numbers is for AD 1708 Sep~13,
where no observers have observed ({\em alldata}), while HS98 give 15 as mean daily number ({\rm dailyrg.dat});
even when using the data in {\em filldata} from HS98 for AD 1708 Sep 13, we obtain 10.3.
The values by HS98 and our values are, however, consistent within $1\sigma$ error bars.
(b) Daily mean after including Becker (with $k^{\prime}=1.11$).
(c) HS98 give $0.0$ here, even though Muller would have never observed
in that month (according to HS98), so that they should assign a $-99.0$ (i.e. n/o or n/a) 
for him for that month meaning that no data are available.
(d) The number 2 (HS98) may be dubious here, because all (4) other observers in HS98 reported one
group, and Becker reported one group with two spots.
(e) G. Kirch would not have observed on AD 1708 Sep 13 \& 14 according to HS98, 
but we found his text about those spots in his letter to Wurzelbaur;
we did take into account 
these two spot observations for re-estimating the group
numbers given in the column {\em mean w/o Becker by us};
the monthly mean for G. Kirch remains to be 1.0.
(f) A number given in brackets (from Becker) is from a different anonymous observer known
to be located in Berlin, which could be G.~Kirch, so that we cannot count that spot independently.
\end{table*}

The sunspot observed by Becker on 1708 Sep~10 and 11 was also observed by La Hire 
on both days, and on one of the two days also by Manfredi, G.~Kirch, and Blanchini, 
while Wideburg says two groups on both days (HS98 and our Tab.~\ref{groupnumbers}).
The spot group (made up of three spots according to Becker) on 1708 Dec~1 was also 
observed by La Hire, Wideburg, and G.~Kirch.
The spot group detected (and drawn) by Becker on 1709 Jan~5--9 was also
noticed by La Hire, Wideburg, G.~Kirch, Wolf, and Muller,
but only on some of the days:
on Jan~5, only Becker saw the two spots forming one group
(no other observer observed on that day, HS98).
Also the spot group found by Becker on 1710 Sep~10
is completely new and was not nocticed before, i.e. not observed
by any other (known) observer; 
La Hire observed from 1710 Sep 4 to 16, but did not detect any spots (HS98) --
but maybe he had just obtained transit observations of the Sun, i.e. did not
search for spots (like e.g. Hevelius, see Carrasco et al. 2015).

We would also like to note that, according to HS98, Wolf detected two groups on AD 1709 Jan~6,
while all other observers detected only one group; since Becker mentioned and drew explicitely, 
that there were two spots forming a group, it is well possible that the {\em two groups} listed
for Wolf are only those {\em two spots} forming one group.

HS98 have assigned correction factors (for the group numbers) for all observers in a similar
manner as Wolf did for sunspot numbers (HS98). Those correction factors are also listed
in Tab.~\ref{groupnumbers}. Since Becker was not known to HS98 as
observer, he does not have a correction factor.
As seen in Tab.~\ref{groupnumbers}, Becker compares very well to 
La Hire (five to six common days with identical numbers and a difference of only 1 on one more day),
Manfredi (two common days, one of them with identical number),
G.~Kirch (six common days with identical numbers on five days and a difference of only 1 on the remaining day), 
Blanchini (with one common day with the identical number),
and Muller (four to five common days, all with identical numbers).
Becker does not compare so well with Wideburg with eight common days, but a difference of $\pm 1$ on three days,
and it is hard to copmpare him with Wolf, as there is only one common day, where the value for Wolf is dubious
(see Tab. 3).
Averaging the correction factors of those five observers with good comparison to Becker
(La Hire, Manfredi, G.~Kirch, Blanchini, and Muller), we could assign a
correction factor of $\sim 1.11 \pm 0.28$ for Becker. According to HS98, 
a restriction to observers with correction factor between 0.6 and 1.4
is applied only to observations since 1848.

In Tab.~\ref{groupnumbers}, we list also the daily, monthly, 
and yearly group sunspot numbers from HS98, and, since 
they are not straight-forward to be reproduced from the raw 
data, also our values derived without using Becker's data,
taking into account the correction factor(s) using equation~2 
in HS98, given here also in Eq.~2. In the last column, we list 
the new values including Becker's observations, as calculated by us. 

The largest change in a monthly or yearly sunspot number, 
compared to HS98, due to additional observations by Becker,
is for January 1709 from 4.3 (HS98) to 4.8 (new), and for 
September 1710 from 0.0 (HS98) to 0.2 (new); the other changes 
are only by $\pm 0.1$ or zero.

\section{Summary}

We found new sunspot observations by Peter Becker (Fig.~1) from Rostock, Germany, 
for the years AD 1708, 1709, and 1710 (Tab.~1), i.e. towards the end of the Maunder Minimum.
We also mention two new sunspot observations by G. Kirch for AD 1708 Sep 13 \& 14
(more newly found observations from G. Kirch will be presented elsewhere).

For AD 1709 Jan~5--10, Becker also provided very good drawings of the two spots observed (Fig.~\ref{1709_01_05}),
so that we could measure their heliographic positions and estimate their areas (Tab.~2).

While most sunspot numbers are consistent with other observers 
and their daily numbers (HS98), there are stronger 
differences on three days: on 1709 Jan~5, only Becker saw the 
two spots forming one group (no other observer observed on 
that day, HS98). Also the spot group found by Becker on 1710 
Sep~10 is completely new and was not nocticed before. La Hire 
did observe, but did not detect any spots (HS98). Then, there 
was previously no observation known for 1709 Jan~3 (HS98), but 
Becker explicitly mentioned that he observed and that there was 
no spot on the Sun on that day.

These newly found old observations change the 
daily, monthly, and yearly sunspot 
group numbers -- compared to HS98 -- slightly, 
in particular for Jan 1709 and Sep 1710 (Tab.~\ref{groupnumbers}),
and there is a small change for the yearly number for 1709.

Since the ``new'' old spots found are not located in the northern 
hemisphere nor at high latitude, they do not indicate the start 
of a new Schwabe cycle, so that the scenario and timing for the
end of the Maunder Grand Minimum around 1712--1715 does not change.

\acknowledgements
We would like to thank the staff at the university archives and libraries at U Rostock and U Schwerin
as well as at the state archive at Rostock. 
RN also acknowledges Hartmut Gilbert (U Jena) for pointing him to the publication by EP
about Becker's sunspots and Andreas Becker (U Rostock) for his help is accessing material
at the U Rostock archive and library.
We are grateful to Andr\'e Kn\"ofel, Richard A\ss{}mann Observatory
Lindenberg, for naming the asteroid 365739 Peterbecker.
We consulted the data base of Hoyt \& Schatten (1998) at
ftp://ftp.ngdc.noaa.gov/STP/space-weather/solar-data/solar-indices/sunspot-numbers/group;
we took the ephemeris for the heliographic coordinates of the solar disk 
from the JPL Horizons ephemeris 
service at http://ssd.jpl.nasa.gov/horizons.cgi.

{}

\end{document}